# A pilot experience in physics laboratory for a vocational school

Vera Montalbano, Maria De Nicola, Simone Di Renzone and Serena Frati

Department of Physical Sciences, Earth and Environment, University of Siena, Siena, Italy

Abstract – date of last change of the paper: 13.04.2014

*The reform of the upper secondary school in Italy has recently introduced physics in the curricula of professional schools, where it was absent. Many teachers, often with a temporary position, are obliged to teaching physics in schools where the absence of the laboratory is added to the lack of interest of students who feel this matter as very far from their personal interests and from the preparation for the work which could expect from a professional school. We report a learning path for introducing students to the measurement of simple physical quantities, which continued with the study of some properties of matter (volume, mass, density) and ending with some elements of thermodynamics. Educational materials designed in order to involve students in an active learning, actions performed for improving the quality of laboratory experience and difficulties encountered are presented. Finally, we compare the active engagement of these students with a similar experience performed in a very different vocational school.*

## Introduction

In few years the Italian educational system has been profoundly reformed. Secondary school reform introduced physics in the curricula of all vocational and technical schools, without any real involvement of teachers that work in these educational situations. Many teachers are facing realities in which the laboratory is absent and the students feel the matter as something useless for their professional training. This innovation may have been inspired by reflections on professional education referring to other realities [1, 2] but it has not been accompanied with neither a specific training for teachers nor funds for implementing laboratories

In the first year of the reform, one of the authors (M. D. N.) was teaching physics at a vocational school for commercial and tourist services where the new curriculum has included physics for 2 hours weekly only in the first class. The school was without laboratories but very close to the university.

Thus, she decided to use the support provided by the National Plan for Science Degree [3-5] (Piano Nazionale Lauree Scientifiche, i. e. PLS) to physics teachers without or with poor physics laboratory. The project was designed like a training activity for qualified teachers in a course of Master in Physics Educational Innovation and Orienting (IDIFO3) [6, 7] and realized within the PLS.

The National Plan for Science Degree is promoted by the Ministry of Education and Scientific Research with the main purpose of contrasting the decline of students' interest in learning physics and the consequent decrease of enrolments science degree in Italy. The main actions of PLS has been to promote professional development for teachers and to orient students essentially by means of laboratory activities. The student is considered the main character of learning, laboratory is mainly a method more than a place and joint planning by teachers and university is encouraged. In the last years, PLS have supported teachers in tackling difficulties encountered in the implementation of the various reforms which have been introduced in secondary school.

## The Project

Since students were not schooled and poorly motivated, we decided in favour of designing a course of introduction to the measurement of simple physical quantities (length and area) in classroom. Then, the study of some properties of matter, such as volume, mass, density, continued in laboratory. The next step was to introduce some elements of thermodynamics (from the concepts of everyday use to the operational definition of temperature and heat). The final scheduled task was the construction and calibration of a thermometer.

The learning path was provided for two 1st classes (another class was excluded because it was considered by their teacher too undisciplined for a correct and save participation) for a total of about 50 students (ages 14-15).

For the activities in the lab, always carried out during school hours, we scheduled 3 experiences (3 hours each one).

All the materials used in the laboratory were easily available and inexpensive (fig. 1) to encourage the school to become independent in the near future.

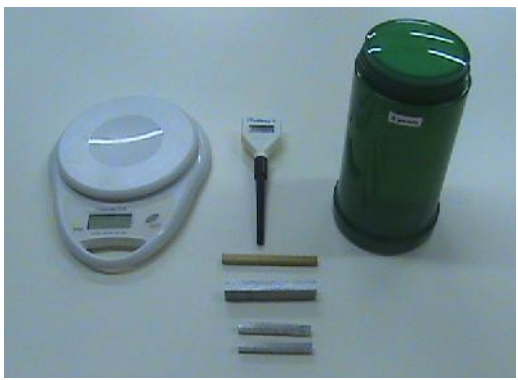

Figure 1. A balance, a digital thermometer, a Dewar vessel and various metal bars were utilized for investigating thermal equilibrium.

A special care was paid in designing the operations to be done in order to ensure the safety of everyone even in the case of students not well accustomed to the correct behavior in laboratory.

## In Laboratory

They were divided in small groups (3 or 4 students) and supported by at least 3 teachers (a university lecturer and two teachers). The first lab was dedicated to direct and indirect measurements of physical quantities, such as mass, volumes and densities. In the second lab, students started by observing the thermal equilibrium between solids (same mass and material, different mass and same material, same mass and different material) and solid/water in a Dewar flask. Then, a measure of heat capacity was realized.

The third lab should have been dedicated to construction and calibration of a thermometer but it was not realized for the reasons described in the next section.

Figure 2. The worksheet contains simple operative explanations, all formulas, hints for evaluating and reducing uncertainties, spaces where to write measures and calculations.

The organization of the laboratory was focused on maintaining students engaged on operational aspects through a detailed worksheet (see fig. 2) carefully designed jointly with teachers that had a very concrete idea of the needs of their students [see the complete worksheets in ref. 8 for more details]. Hints on measurements, calculations for analyzing and comparing the data were given. A particular emphasis was paid on safety aspects (e.g. they obtained the boiling water directly into the Dewar vessel). Finally, the experience was discussed again in the classroom, the calculations completed and the conclusions drawn.

Each student had to return his personal worksheet which was assessed by the teacher.

## Discussion and Conclusions

The initial effort to improve the teaching of physics in a vocational school by adapting a introductory learning path, well tested in other schools, on measurements of simple physical quantities and some elements of thermodynamics led to an unsatisfactory result. However, it can allow to understand why it is often difficult to extend the results of physics education research into classroom practice.

This pilot study started by selecting two classes on the three potentially interested ones and none of them realized the full learning path. Despite the care and attention paid in designing educational material and performing the activities in the lab, the learning process was not comparable to that obtained with students in other types of upper secondary schools or even with younger students in lower secondary school.

Many critical aspects emerged in the realization of the learning path:

- Management of activity outside the school, vacation and other difficulties delayed the labs, the two experiences were carried out only one month one from each other.

- Labs were carried out at the end of the school year, over a period not well suited.

- Students had not enough time for completing activities, which confirms the need to dilute the work in more parts, by planning less activities in each lab session.

- Dead times during the experimentation must be avoid, because pupils can have a decreasing of attention.

- Management of the security aspects can be more effective from educational point of view.

- Strategies are needed to reinforce the group's collaboration in the production of the common work, to avoid the tendency that the work is done by few.

The average level of the class and previous remarks led to mediocre results, for this reason the teacher gave a particular value in the assessment to participation, care and operative skills.

Students' behavior in laboratory observed through all teachers' observations, their informal comments and the examination of all other conditions that influenced the teaching/learning process were carefully analyzed and discussed in order to identify the main obstacles to the learning process. Mainly, it is possible to recognize different kinds of trouble source:

- The lack of motivation and interest is a real problem in this context, the hands-on activities are always appreciated but this is not enough if the topic is perceived as useless. Looking for examples more connected to their future professional world could favour the learning process.

- External conditions can be determinant for achieving acceptable results or an overwhelming defeat. In particular, the availability of a laboratory with a minimum of materials and a technical support that could allow to realize hands-on activity in a more effective way can make the difference in this kind of school. Thus, it could be possible to develop students skills in laboratory through less intense sessions in which they can have more time for inquiring, exploring, reflecting and consolidate their learning.

- A more effective organization in lab designing is necessary in this case, each step in the use of spaces and times must be carefully planned, tested and adapted to students reactions. In particular, the dynamics in and between groups must be investigated more in these borderline context in order to develop strategies useful in physics education.

Overall, students were involved in first steps of an active learning process, completely absent in other lectures, but the achieved results were still partial and disappointing from the point of view of physics learning. At the end, a consistent number of students used correctly the unit of measure, but not all. Almost all are able to evaluate the instrumental uncertainty for a measure in very easy cases, but few apply correctly the propagation of errors in an indirect measure. Finally, very few (not more than two or three) acquired confidence with calorimetric measurements or were able to compare correctly two measures. These results confirm that laboratory is a powerful tool for learning physics but more research in physics education is necessary for rendering effective learning processes in this difficult context.

Moreover, a first important step is to obtain the physics laboratory at school. Another important point is that physics teachers need more support in this case. The pilot experience was performed two years ago and there was no way to continue this experience because all teachers try to stay less than possible in this school so every few months they change. Laboratory is still lacking and there is no sign of new projects for improving the learning process in physics such as in all other relevant but inadequate matters, e. g. mathematics.

 The understanding of which aspects can be improved in this pilot experience has been an essential step for designing a successive learning path [9] in a different vocational school (an agricultural technical institute). In this case, the aim was to make interesting measures of physical quantities, calorimetry and state transitions connecting them to an instrument that students use in their professional career. Second classes has been involved and the main purpose of increasing students' attention on physics has been fully achieved [9].

## References

[1] Guile, D., & Young, M., Transfer and transition in vocational education: Some theoretical considerations, in T. Tuomi-Gröhn & Y. Engertröm (Eds.), Between school and work. New perspectives on transfer and boundary-crossing (pp. 63-81), Oxford. UK: Elsevier Science (2003).

[2]  Shaping the Future 3 Physics in Vocational Courses, Edited  by  Ken Gadd. Bristol: Institute of  Physics Publishing (1999), ISBN 0 7503 06831.

[3]   PLS   Website,   (in Italian).   [online]. [cit. 30. 6. 2013].   Available from: www.progettolaureescientifiche.eu

[4] Chiefari G., Sassi E., Testa I. (2012), Improving secondary students' scientific literacy and  laboratory skills: the Italian Project "Scientific Degrees", in  Proceedings of The World Conference on Physics Education 2012, editor Mehmet Fatih Ta ar,  p. 53-63, Ankara: Pegem Akademi (2014). [online]. [cit. 31. 1. 2014]. Available from: http://www.wcpe2012.org/.

[5]  Montalbano V., Fostering Student Enrollment in Basic Sciences: the Case of Southern Tuscany, in  Proceedings of The 3rd International Multi-Conference on Complexity, Informatics and Cybernetics: IMCIC 2012, ed. N. Callaos et al, 279, (2012), arXiv:1206.4261 [physics.ed-ph], [online]. [cit. 4. 10. 2013]. Available from: http://arxiv.org/abs/1206.426.

[6] Francaviglia M., Lorenzi M. G., Michelini M., Santi L., Stefanel A.,  IDIFO3 - teachers formation on  modern physics  and mathematical foundations  of quantum physics: a cross  sectional  approach, J. Appl. Math. 5, No. 1(2012), p. 231- 240

[7]  Battaglia R., Cazzaniga L., Corni F., De Ambrosis A., Fazio C., Giliberti M., Levrini O. , Michelini M., Mossenta A., Santi L., Sperandeo R. M., Stefanel A.,Master IDIFO (Innovazione Didattica in Fisica e Orientamento): a community of  italian physics

education researchers and teachers as a model for a research based in-service teacher development in modern physics, in D. Raine et al. eds., Physics Community And Cooperation Volume 2, GIREP-EPEC & PHEC 2009 International Conference. p. 97-136, LEICESTER:Lulu / The Centre for Interdisciplinary Science, ISBN: 978-1-4466-1139-5.

[8] Lab worksheets (in italian) [online]. [cit. 11. 4. 2014]. Available from: https://www.researchgate.net/publication/261511275_scheda_masse_densit_volumi?ev=prf_pub , https://www.researchgate.net/publication/261511281_Equilibrio_termico?ev=prf_pub.

[9] Benedetti R., Mariotti E., Montalbano V., A First Steps into Physics in the Winery, in Proceedings of The World Conference on Physics Education 2012, editor Mehmet Fatih Ta ar, p. 293-301, Ankara: Pegem Akademi (2014). [online]. [cit. 31. 1. 2014]. Available from: http://www.wcpe2012.org/.